\documentclass[twocolumn,showpacs,superscriptaddress,preprintnumbers,amsmath,amssymb,prb]{revtex4-1}

\usepackage{graphicx}
\usepackage{dcolumn}
\usepackage{bm}
\usepackage{color}

\begin{document}


\title{A systematic study of magnetodynamic properties at finite temperatures in doped permalloy from first principles calculations }

\author{Fan Pan}
\email{fanpan@kth.se}
\affiliation{Department of Materials and Nano Physics, School of Information and Communication Technology, KTH Royal Institute of Technology, Electrum 229, SE-16440 Kista, Sweden}
\affiliation{Swedish e-Science Research Center (SeRC), KTH Royal Institute of Technology, SE-10044 Stockholm, Sweden}

\author{Jonathan Chico}
\affiliation{Department of Physics and Astronomy, Materials Theory Division, Uppsala University, Box 516, SE-75120 Uppsala, Sweden}

\author{Johan Hellsvik}
\affiliation{Department of Materials and Nano Physics, School of Information and Communication Technology, KTH Royal Institute of Technology, Electrum 229, SE-16440 Kista, Sweden}

\author{Anna Delin}
\affiliation{Department of Materials and Nano Physics, School of Information and Communication Technology, KTH Royal Institute of Technology, Electrum 229, SE-16440 Kista, Sweden}
\affiliation{Swedish e-Science Research Center (SeRC), KTH Royal Institute of Technology, SE-10044 Stockholm, Sweden}
\affiliation{Department of Physics and Astronomy, Materials Theory Division, Uppsala University, Box 516, SE-75120 Uppsala, Sweden}

\author{Anders Bergman}
\affiliation{Department of Physics and Astronomy, Materials Theory Division, Uppsala University, Box 516, SE-75120 Uppsala, Sweden}

\author{Lars Bergqvist}
 \affiliation{Department of Materials and Nano Physics, School of Information and Communication Technology, KTH Royal Institute of Technology, Electrum 229, SE-16440 Kista, Sweden}
 \affiliation{Swedish e-Science Research Center (SeRC), KTH Royal Institute of Technology, SE-10044 Stockholm, Sweden}

\date{\today}

\begin{abstract}
By means of first principles calculations, we have systematically investigated how the magnetodynamic properties Gilbert damping, magnetization and exchange stiffness are affected when permalloy (Py) (Fe$_{0.19}$Ni$_{0.81}$) is doped with 4d or 5d transition metal impurities. We find that the trends in the Gilbert damping can be understood from relatively few basic parameters such as the density of states at the Fermi level, the spin-orbit coupling and the impurity concentration. 
The temperature dependence of the Gilbert damping is found to be very weak which we relate to the lack of intraband transitions in alloys.
Doping with $4d$ elements has no major impact on the studied Gilbert damping, apart from diluting the host. However, the $5d$ elements have a profound effect on the damping and allows it to be tuned over a large interval while maintaining the magnetization and exchange stiffness.
As regards spin stiffness, doping with early transition metals results in considerable softening, whereas late transition metals have a minor impact.
Our result agree well with earlier calculations where available. In comparison to experiments, the computed Gilbert damping appears slightly underestimated while the spin stiffness  show good general agreement. 
%


\end{abstract}

\maketitle

\section{\label{sec:intro}Introduction}

%
%


Spintronics and magnonic applications have attracted a large degree of attention due to the potential of creating devices with reduced energy consumption and improved performance compared to traditional semiconductor devices \cite{Igor2004,Brataas2012,Gomonay2014}.
An important ingredient for understanding and improving the performance of these devices is a good knowledge of the magnetic properties. In this study, we focus on the saturation magnetization $M_s$, the exchange stiffness $A$ and the Gilbert damping $\alpha$ \cite{llgequation}. The latter is related to the energy dissipation rate of which a magnetic system returns to its equilibrium state from an excited state, e.g. after the system has been subjected to an external stimulii such as an electrical current which alters its magnetic state. The three parameters, $M_s$, $A$ and $\alpha$ describe the magnetodynamical properties of the system of interest. Ultimately one would like to have complete independent control and tunability of these properties. In this study, the magnetodynamical properties of Permalloy (Py) doped with transition metal impurities are systematically investigated within the same computational framework.

The capability of tuning the damping for a material with such a technological importance as Py is important for the development of possible new devices in spintronics and magnonics.
The understanding of how transition metals or rare earth dopants can affect the properties of Py has been the focus of a number of recent experimental studies~\cite{Rantschler2007,Shigemi2011,Woltersdorf2009,Endo2012,tunable_yuli}.
Typically in these studies, the ferromagnetic resonance \cite{Griffiths1946} (FMR) technique is employed and $\alpha$ and $M_s$ are extracted from the linewidth of the uniform precession mode while $A$ is extracted from the first perpendicular standing spin-wave mode \cite{Seavey1958,Portis1963}. On the theory side, calculations of Gilbert damping from first-principles density functional theory methods have only recently become possible due to the complexity of such calculations. Two main approaches have emerged, the breathing Fermi surface model \cite{Kambersky1970,Fahnle2006} and the torque correlation models \cite{Kambersky1976,Gilmore2007}. Common to both approaches is that spin-orbit coupling along with the density of states at the Fermi level are the main driving forces behind the damping. The breathing Fermi surface model only takes only into account intraband transitions while torque correlation model also includes interband transitions. The torque correlation model in its original form contains a free parameter, namely the scattering relaxation time. Brataas {\it et. al} \cite{ScaThy_Gil_Dp} later lifted this restriction by employing scattering theory and linear response theory. The resulting formalism provides a firm foundation of calculating $\alpha$ quantitative from first principles methods and allows further investigations of the source of damping. Gilbert damping in pure Py as well as doping with selected elements have been calculated in the past\cite{Starikov2010,Ebert2011b,Mankovsky2013,tunable_yuli}, however no systematic study of the magnetodynamic properties within the same computational framework has been conducted which the present paper aims to address. 

The paper is outlined as follows: In Section~\ref{sec:methods} we present the formalism and details of the calculations, in Section~\ref{sec:results} we present the results of our study and in Section~\ref{sec:summary} we summarize our findings and provide an outlook.

\section{\label{sec:methods}Theory}
%
\subsection{Crystal structure of permalloy and treatment of disorder in the first-principles calculations} \label{sec:cs}
Pure Permalloy (Py), an alloy consisting of iron (Fe) and nickel (Ni) with composition Fe$_{0.19}$Ni$_{0.81}$, crystallizes in the face centered cubic (fcc) crystal structure, where Fe and Ni atoms are randomly distributed. Additional doping with $4d$ and $5d$ impurities (M) substitutes Fe (or Ni) so that it becomes a three component alloy with composition Py$_{1-x}$M$_x$, where $x$ is the concentration of the dopant. 

All first principles calculations in this study were performed using the spin polarized relativistic  (SPR) Korringa-Kohn-Rostoker (KKR) \cite{Ebert2011a} Green's function (GF) approach as implemented in the SPR-KKR software\cite{SPRKKR}. The generalized gradient approximation (GGA) \cite{GGA_PBE} was used in the parametrization of the exchange correlation potential and both the core and valence electrons were solved using the fully relativistic Dirac equation. The broken symmetry associated with the chemical substitution in the system was treated using the coherent potential approximation (CPA) \cite{CPA1, CPA2}.

\subsection{\label{sec:Compoa}Calculation of magnetodynamical properties of alloys: Gilbert damping within linear response theory and spin stiffness}
%
One of the merits with the KKR-CPA method is that it has a natural way of incorporating calculations of response properties using linear response formalism \cite{ScaThy_Gil_Dp,Ebert2011b,Mankovsky2013}. 

The formalism for calculating Gilbert damping in the present first principles method has been derived in Refs. [\onlinecite{ScaThy_Gil_Dp}] and [\onlinecite{Mankovsky2013}], here we only give a brief outline of the most important points. The damping can be related as the dissipation rate of the magnetic energy which in turn can be associated to the Landau-Lifshitz-Gilbert (LLG) equation \cite{llgequation}, leading to the expression   

\begin{equation}
\dot{E} = H_{\text{eff}} \cdot \frac{\text{d}\mathbf{M}} {\text{d}\tau}
=\frac{1}{\gamma^2}\dot{\hat{\mathbf{m}}}[\tilde{G}(\mathbf{m})\dot{\hat{\mathbf{m}}}],
\label{eq:E_dm/dt}
\end{equation}
where $\hat{\mathbf{m}}=\mathbf{M}/M_s$ denotes the normalized magnetization vector, $M_s$ the saturation magnetization, $\gamma$ the gyromagnetic ratio and $\tilde{G}(\mathbf{m})$ the Gilbert relaxation rate tensor.


Perturbing a magnetic moment from its equilibrium state by a small deviation, $\hat{\mathbf{m}}(\tau)=\hat{\mathbf{m}}_0+\mathbf{u}(\tau)$, gives an alternative expression of the dissipation rate by employing linear response theory

\begin{equation}
\begin{split}
\label{eq:E_dis}
\dot{E}_{dis} = & \pi \hbar \sum_{ij}\sum_{\mu \nu} \dot{u}_{\mu}\dot{u}_{\nu}\langle\psi_{i} | \frac{\partial\hat{H}}{\partial u_{\mu}} | \psi_{j} \rangle\langle \psi_{j} | \frac{\partial \hat{H}}{\partial u_{\nu}} | \psi_{i} \rangle \times \\
& \delta (E_F -E_i)\delta(E_F-E_j),
\end{split}
\end{equation}
where the $\delta$-functions restrict the summation over eigenstates to the Fermi level which can be rewritten in terms of Green's function as $\text{Im}G^+(E_F)=-\pi\sum_{i}|\psi_i\rangle\langle\psi_i|\delta(E_F-E_i)$. By comparing Eqs.~(\ref{eq:E_dm/dt}) and 
(\ref{eq:E_dis}), the Gilbert damping parameter $\alpha$ is obtained, which is dimensionless and is related to the Gilbert relaxation tensor $\alpha = \tilde{G}/(\gamma M_s)$. This can be expressed as a transport Kubo-Greenwood-like equation \cite{Kubo,Greenwook} in terms of the retarded single-particle Green's functions 
 
\begin{equation}
\alpha_{\mu \nu} = -\frac{\hbar\gamma}{\pi M_s}\text{Trace}\Big{\langle}\frac{\partial \hat{H}}{\partial u_{\mu}}\text{Im}G^{+}(E_F)\frac{\partial \hat{H}}{\partial u_{\nu}}\text{Im}G^{+}(E_F)\Big{\rangle}_c,
\label{eq:alpha_mat}
\end{equation}
where $\langle \dots \rangle_c$ denotes a configurational average. For the cubic systems treated in this study, the tensorial form of the damping can with no loss of generality be replaced with a scalar damping parameter.  Thermal effects from atomic displacements and spin fluctuations were included using the alloy-analogy model \cite{thermal_phonon_magnon} within CPA. 

The spin-wave stiffness $D$ is defined as the curvature of the spin wave dispersion spectrum at small wave vectors ($\omega(\mathbf{q})\approx D \mathbf{q}^2$). $D$ in turn is directly related to the exchange interactions in the Heisenberg model which are obtained using the LKAG formalism \cite{Liechtenstein1987,Ebert2009} such that 





\begin{equation}
\label{eq:D_origin}
D = \frac{2}{3}\sum_{ij} \frac{J_{ij}R_{ij}^2}{\sqrt{m_i m_j}},
\end{equation}
where $J_{ij}$ is the interatomic exchange parameter between the $i$-th and $j$-th magnetic moment, $R_{ij}$ the distance connecting the atomic sites with index $i$ and $j$ and $m_i$ ($m_j$) the magnetic moment at site $i$ ($j$). It is worth noting that Eq.~(\ref{eq:D_origin}) only holds for cubic systems as treated here, for lower symmetries the relation needs modifications.
The exchange couplings in metallic systems are typically long ranged and could have oscillations of ferromagnetic and antiferromagnetic character, such as present in RKKY type interactions. 
Due to the oscillations in exchange interactions, care is needed to reach numerical convergence of the series in Eq.~(\ref{eq:D_origin}) and it is achieved following the methodology as outlined in Refs.~\onlinecite{Pajda} and \onlinecite{Phillip}.

\subsection{\label{sec:spinstiff}Calculation of finite temperature magnetic properties }

%
Once the exchange interactions within the Heisenberg model have been calculated, we obtained finite temperature properties from Metropolis \cite{Mentroplis1953} Monte Carlo simulations as implemented in the UppASD software package \cite{UppASD,UppASD_link}. In particular, the temperature dependent magnetization was obtained, and enters the expression for micromagnetic exchange stiffness A, defined as \cite{calculate_A,Hamrle,aharoni,coey}


\begin{equation}
A(T) = \frac{D M(T)}{2g\mu_B},
\label{spin_stiff_A}
\end{equation}
where $\mu_B$ is the Bohr magneton, $g$ is the Land\'e g-factor and $M(T)$ the magnetization at temperature $T$. 


\subsection{Details of the calculations} \label{sec:MC}
%

For each concentration of the different impurities in Py, the lattice parameter was optimized by varying the volume and finding the energy minimum. The k-point mesh for the self consistent calculations and exchange interactions was set to 22$^3$ giving around 800 k-points in the irreducible wedge of the Brillouin zone (IBZ). The Gilbert damping calculation requires a very fine mesh to resolve all the Fermi surface features and therefore a significantly denser k-point mesh of 228$^3$ ($\sim1.0\times 10^6$ k-points in IBZ) was employed in these calculations to ensure numerical convergence. Moreover, vertex corrections \cite{Butler1985} were included in the damping calculations since it has been revealed to be important in previous studies \cite{Mankovsky2013} for obtaining quantitative results. 

\section{Results} \label{sec:results}

\subsection{Equilibrium volumes and induced magnetic moments}
%

 \begin{figure}[htbp]
\begin{center}
\includegraphics[width=3.4in]{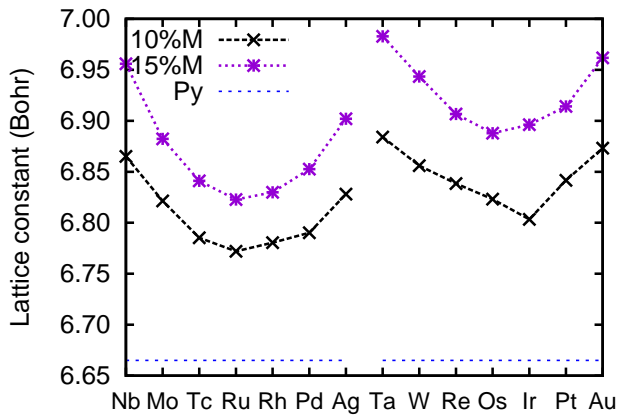}
\end{center}
\caption{\label{fig:lc_gga}
 Calculated equilibrium volumes of Py-M, where M stands for a $4d$ (left) or $5d$ (right) transition metal. Values for $10\%$ and $15\%$ doping concentrations are shown. Reference value of pure Py is diplayed with a dashed line.
 }
\end{figure}

 Figure ~\ref{fig:lc_gga} shows the calculated equilibrium volume of doped Py for two different concentrations ($10\%$ and $15\%$) of impurities from the $4d$ and $5d$ series of the Periodic Table. First of all, it is noted that the volume increases with the concentration, and the volume within a series ($4d$ or $5d$) has a parabolic shape with minimum in the middle of the series. This is expected since bonding states are consecutively filled and maximized in the middle of the series and thus the bonding strength reaches a maximum. Moving further through the series, anti-bonding states start to fill, giving rise to weaker bonding and larger equilibrium volumes. This is consistent with the atomic volumes within the two series \cite{4d_atomic_radius}.


\begin{figure}[htbp]
\begin{center}
\includegraphics[width=3.4in]{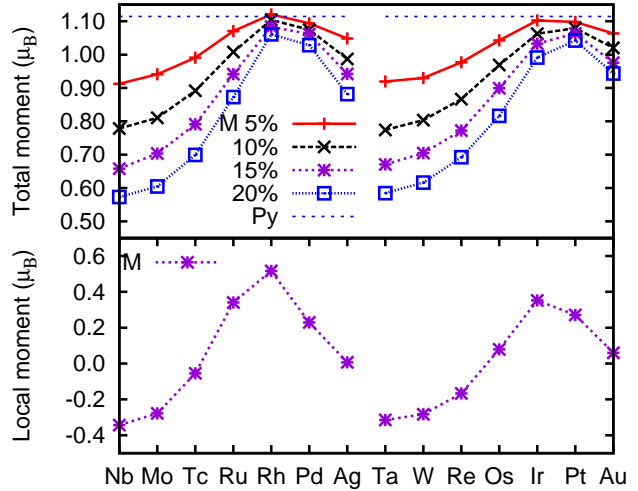}
\end{center}
\caption{\label{fig:multi_moments}
(Upper) Total magnetic moment (spin and orbital) for different impurities and concentrations. Reference value for pure Py marked with a dashed line. (Lower)
Local impurity magnetic moment for Py$_{0.95}$M$_{0.05}$}
\end{figure}

The local moments of the host atoms are only weakly dependent on the type of impurity atom present. Moreover, the magnetic moments are dominated by the spin moment $\mu_S$ while the orbital moments $\mu_L$ are much smaller. As an example, in pure Py without additional doping, the spin (orbital) moments of Fe is calculated to $\approx$ 2.64 (0.05) $\mu_B$ and for Ni $\approx$ 0.64 (0.05) $\mu_B$, respectively. This adds up to an average spin (orbital) moment of $\approx$ 1.04 (0.05) $\mu_B$ by taking into account the concentration of Fe and Ni in Py. The total moment is analyzed in more detail in
Fig.~\ref{fig:multi_moments} (upper panel). As mentioned above, one would like to achieve tunable and independent control of the saturation magnetization. Reducing the magnetization reduces the radiative extrinsic damping but could at the same time affect the other properties in an unwanted manner. In many situations, one strive for keeping the value of the total moment (saturation magnetization) at least similar to pure Py, even for the doped systems. It is immediately clear from Fig.~\ref{fig:multi_moments} that doping elements late in the series are the most preferable in that respect, for instance Rh and Pd in the $4d$ series and Ir, Pt and Au in the $5d$ series.

In Fig. \ref{fig:multi_moments} (lower panel) we show the local impurity magnetic moments for $5\%$ impurities in Py. In the beginning of the $4d$ ($5d$) series, the impurity atoms have an antiferromagnetic coupling, reflected in the negative moments compared to the host (Fe and Ni) atoms while latter in the series couples ferromagnetically (positive moments). The antiferromagnetic coupling may not be preferred since it will tend to soften the magnetic properties and maybe even cause more complicated non-collinear magnetic configurations to occur.

\subsection{Band structure}
\begin{figure}[htbp]
\begin{center}
\includegraphics[width=3.4in]{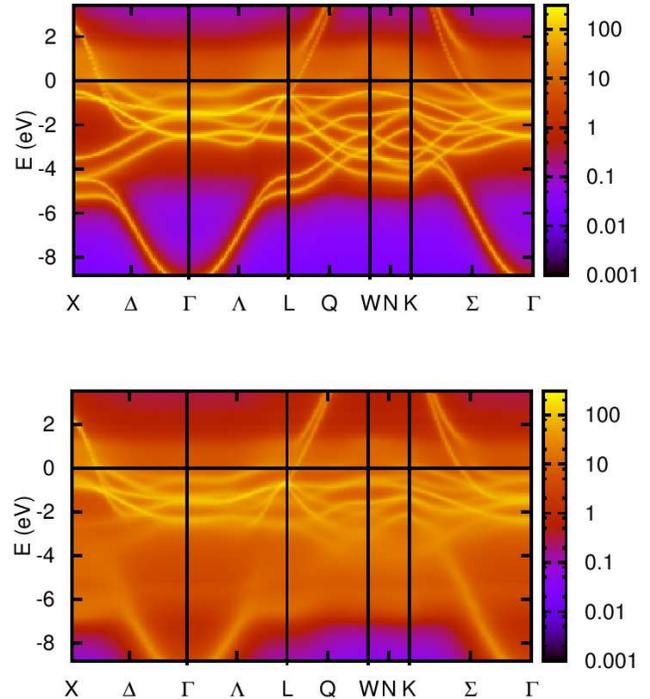}
\end{center}
\caption{\label{fig:Py_Py20Pt_bsf}
The Bloch spectral function $A(E,\mathbf{k})$ of Py (upper panel) and Py doped with $20\%$ Pt impurities (lower panel). The Fermi level is indicated with a horizontal black line at zero energy.}
\end{figure}

Since Py and doped-Py are random alloys, they lack translational symmetry and calculations using normal band structure methods are more challenging due to the need for large supercells. However, employing CPA restores the translational symmetry and more importantly, the band structure of disordered systems can be analyzed through the Bloch spectral function (BSF) $A(E,\mathbf{k})$, which can be seen as a wave vector $\mathbf{k}$-dependent density of states (DOS) function.  For ordered systems the BSF is a $\delta$-like function at energy E($\mathbf{k})$ while for disordered systems each peak has an associated broadening with a linewidth proportional to the amount of disorder scattering. In the upper panel of  Fig.~\ref{fig:Py_Py20Pt_bsf} the calculated BSF for pure Py is displayed. Despite being a disordered system, the electron bands are rather sharp below the Fermi level while in the vicinity of the Fermi level the bands becomes much more diffuse indicating that most of the disorder scattering takes place around these energies. 

If Py is doped with 20\% Pt impurities, the positions of the electron bands do not change much as shown by the BSF in the lower panel of Fig.~\ref{fig:Py_Py20Pt_bsf}. The most striking change is the large increase of the disorder scattering compared to than Py causing diffuse electron bands throughout the Brillouin zone and energies. However, exactly at the Fermi level the differences between the doped and undoped system is not very pronounced and these states are the most important for the determination of the Gilbert damping, as seen from Eq.~\ref{eq:alpha_mat}.

\subsection{
Gilbert damping: effect of doping}
%

\begin{figure}[htbp]
\begin{center}
\includegraphics[width=3.4in]{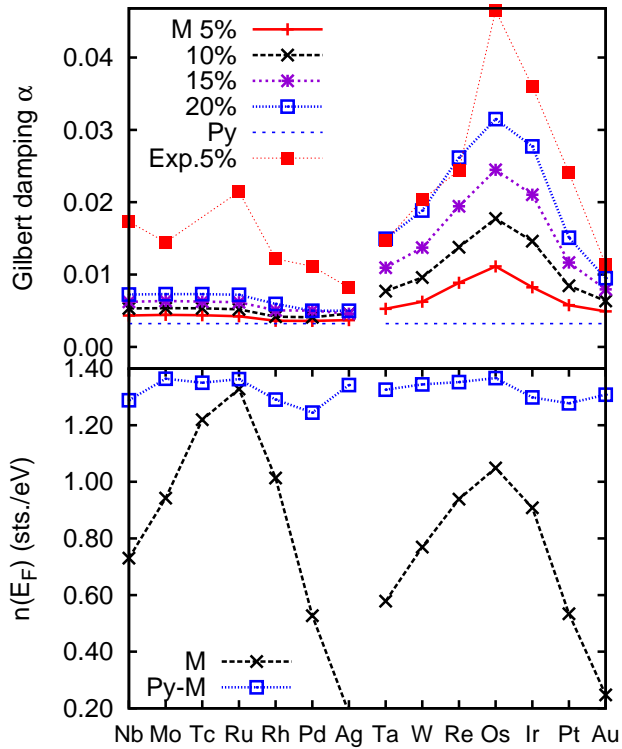}
\end{center}
\caption{\label{fig:dos_damping_all_con}
(Upper) Calculated Gilbert damping parameter for Py+M in different concentrations of $4d$ and $5d$ transition metal M at low temperatures ($T=10\text{K}$). Experimental results from Ref.~[\onlinecite{Rantschler2007}] measured at room temperature are displayed by solid squares and dashed line indicate reference value for pure Py. (Lower) Total (blue) and impurity (black) density of states at the Fermi level $E_F$ for 10\% impurities in Py. 
}
\end{figure}

The calculated Gilbert damping of the doped Py systems for different concentrations of impurities is shown in  Fig.~\ref{fig:dos_damping_all_con} (upper panel). The $4d$ impurities only marginally influence the damping while the $5d$ impurities dramatically change the damping. The first observation is that we obtain very good agreement as in the previous study\cite{Ebert2011b} for the $5d$ series with 10\% impurities, however not so surprising since we use same methodology. Secondly, the most dramatic effect on damping upon doping is for the case of Py doped with 20\% Os impurities in which the damping increases with approximately 800\% compared to pure Py, as previously reported in Ref.~[\onlinecite{Mankovsky2013}]. However, in the present study we have systematically varied the impurity elements and concentrations and tried to identify trends over a large interval. Compared to experiments\cite{Rantschler2007}, the calculated values of the Gilbert damping are consistently underestimated. However it is worth remembering that calculations only shows the intrinsic part of the damping while experiments may still have some additional portion of extrinsic damping left such as Eddy current damping and radiation damping, since it is difficult to fully separate the different contributions. Moreover, in calculations a complete random distribution of atoms is assumed while there may be sample inhomogeneities such as clustering in the real samples.

From most theoretical models, the two main material properties that determine the damping are the density of states (DOS) at the Fermi level and the strength of the spin-orbit coupling. In the following, we first investigate separately how these properties affect the damping and later the combination of the two. In the lower panel of Fig.~\ref{fig:dos_damping_all_con}, the total DOS and the impurity-DOS are displayed for 10\% impurity concentration of $4d$ and $5d$ series transition metals. In the both $4d$ and $5d$ series the impurity-DOS exhibits a maximum in the middle of the series. However, the value of the DOS are similar for the $4d$ and $5d$ series and  therefore cannot solely explain the large difference in damping found between the two series. For the $4d$ series, the calculated damping is not directly proportional to the DOS while there is a significant correlation of the DOS and damping in the $5d$ series. 

\begin{figure}[htbp]
\begin{center}
\includegraphics[width=3.4in]{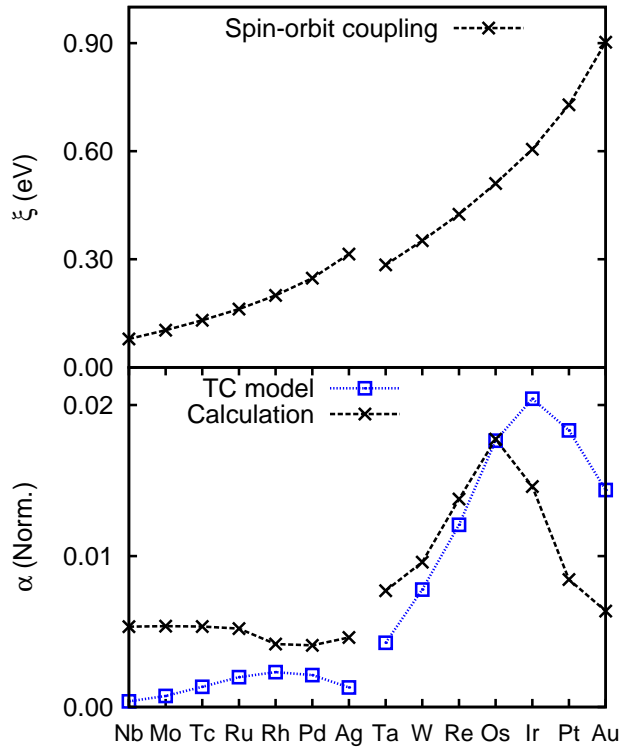}
\end{center}
\caption{\label{fig:soc_parameter}
Upper: the spin-orbit parameter of $d$-electrons of the impurity atoms. Lower: qualitative comparison between calculations and torque correlation (TC) model for damping with 10\% impurity concentration. 
}
\end{figure}

In order to analyze the separate influence of spin-orbit coupling on the damping, we show in upper panel of Fig.~\ref{fig:soc_parameter} the spin-orbit parameter $\xi \propto \frac{1}{r} \frac{dV(r)}{dr}$, where V(r) is the radial potential, of the impurity $d$-states. The calculations include all relativistic effects by solving the Dirac equation but here we have specifically extracted the main contribution from the spin-orbit coupling. As expected, the spin-orbit parameter increases with atomic number $Z$, and is therefore considerably larger in the $5d$ series compared to the $4d$ series. This is the most likely explanation why the damping is found to be larger in the $5d$ series than the $4d$ series. However, within a single element in either the 4d or 5d series, the damping is quadratically dependent on the relatve strength of the spin-orbit strength\cite{Mankovsky2013}. 
The calculated values of the spin-orbit parameter are in good agreement with previous calculations \cite{Popescu2001,Christensen1984} and reaches large values of 0.6-0.9 eV for the late $5d$ elements Ir,Pt and Au while all values are below 0.3 eV for the $4d$ series. If the damping across elements would only be proportional to the spin-orbit coupling, then the damping would monotonously increase with atomic number and since this is not what happens, we conclude that there is a delicate balance between spin-orbit coupling and DOS that determines the damping which is further highlighted through a qualitative analysis of the involved scattering processes. 

In the torque correlation model, the dominant contribution to damping is through the scattering \cite{Kambersky2007,Farle} and takes the following form

\begin{equation} \label{eq:scatt}
\alpha =\frac{1}{\gamma M_s} (\frac{\gamma}{2})^2 n(E_F) \xi^2 (g-2)^2/\tau,
\end{equation}
where $\tau$ is the relaxation time between scattering events, and $g$ the Lande g-factor, for small orbital contributions, can be related as \cite{Meyer1961} $g=2(1+\frac{\mu_L}{\mu_S}$). We assume that $\tau$ is the same for all impurities, which is clearly an approximation but calculating $\tau$ is beyond the scope of the present study. By normalizing the damping from Eq.~(\ref{eq:scatt}) such that the value for Os (10\% concentration) coincides with the first principles calculations, we obtain a qualitatively comparison between the model and calculations, as illustrated in lower panel of Fig.~\ref{fig:soc_parameter}. It confirms the trend in which $5d$ series lead to a larger damping than the $4d$ series and captures qualitatively the main features. However, the peak value of the damping within the $5d$ series in the TC model occurs for Ir while calculations give Os as in experiment. Another model developed for low dimensional magnetic systems such as adatoms and clusters suggests that the damping is proportional to the product of majority and minority density of states at the Fermi level\cite{Lounis2015}. It produces a parabolic trend but with maximum at incorrect position and fails to capture the increased damping of the $5d$ elements.

\begin{figure}[htbp]
\begin{center}
\includegraphics[width=3.4in]{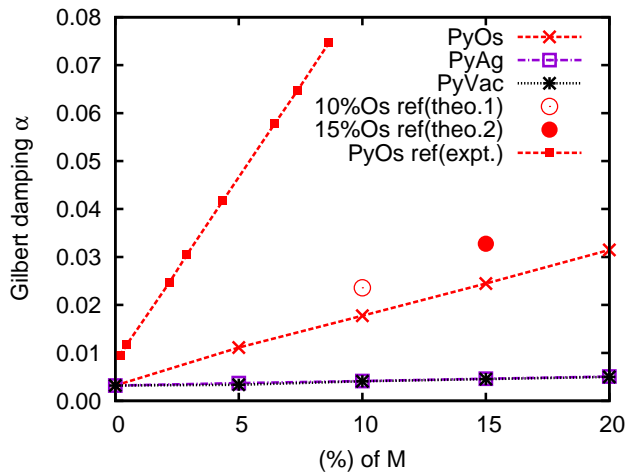}
\end{center}
\caption{\label{fig:damping_Ru_Os_Vac}
Calculated Gilbert damping as a function of Os, Ag and vacancy (Vac) concentration in Py. Open red circle: calculation from Ref.~[\onlinecite{Ebert2011b}], solid red circle: calculation from Ref.~[\onlinecite{Mankovsky2013}] and red solid square: experimental data from Ref.~[\onlinecite{Rantschler2007}]}
\end{figure}

To further analyse the role of impurity atoms on the damping we also performed calculations where instead of impurities we added vacancies in the system, i.e. void atoms. The results are shown in Fig.~\ref{fig:damping_Ru_Os_Vac} where damping as a function of concentration of Ag ($4d$), Os ($5d$) and vacancies are compared to each other along with Os results from experimental \cite{Rantschler2007} and previous calculations. Surprisingly, vacancies have more or less the same effect as Ag with the damping practically constant when increasing concentration. Since Ag has a zero moment, small spin-orbit coupling and small density of states at the Fermi level, the net effect of Ag from a damping (or scattering) point of view is mainly diluting the host similar to adding vacancies. In contrast, in the Os case, being a $5d$ metal, there is a strong dependence on the concentration that was previously analyzed in terms of density of states and Os having a strong spin-orbit coupling. Our results from Os is slightly lower than the previous reported values\cite{Ebert2011b,Mankovsky2013}, despite using same software. However, the most likely reason for the small discrepancy is the use of different exchange-correlation potentials in the two cases.


\subsection{Gilbert damping: effect of temperature}
%
 In the previous section we studied how the damping depends on the electronic structure and spin-orbit coupling at low temperatures. However, with increasing temperature additional scattering mechanisms contribute to the damping, most importantly phonon and magnon scattering. The phonon scattering is indirectly taken into account by including a number of independent atomic displacements bringing the atoms out from their equilibrium positions and magnon scattering is indirectly included by reducing the magnetic moment for a few configurations and then average over all atomic and magnetic configurations within CPA. It should be noted that the present methodology using the alloy-analogy model \cite{thermal_phonon_magnon} has limitations for pure systems at very low temperatures where the damping diverges, but we are far from that situation in this study since all systems have intrinsic chemical disorder. However, the limitations for pure systems can be lifted using a more advanced treatment using explicit calculation of the dynamical susceptibility\cite{Costa2015}.

\begin{figure}[htbp]
\begin{center}
\includegraphics[width=3.4in]{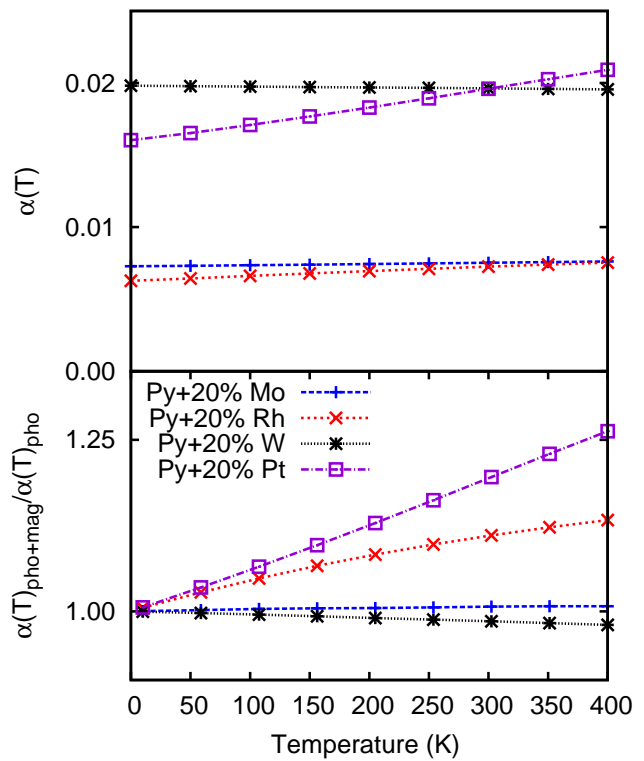}
\end{center}
\caption{\label{fig:damping_pho+mag_ratio}
Gilbert damping parameter including temperature effects from both atomic displacements and spin fluctuations (upper panel). The effect of spin fluctuations on the Gilbert damping (lower panel), see text.
}
\end{figure}

The temperature dependence of damping for a few selected systems is displayed in Fig.~\ref{fig:damping_pho+mag_ratio} where both atomic displacements and spin fluctuations are taken into account. From the $4d$ ($5d$) series, we choose to show results for Mo and Rh (W and Pt), where Mo (W) has a small antiferromagnetic moment and Rh (Pt) a sizeable ferromagnetic moment, from Fig.~\ref{fig:multi_moments}. All systems display an overall weak temperature dependence on damping which only marginally increases with temperature, as shown in upper panel of Fig.~\ref{fig:damping_pho+mag_ratio}. However, in order to separate the temperature contributions from atomic displacements and spin fluctuations, we show the ratio between the total damping and damping where only atomic displacements are taken into account in the lower panel of Fig.~\ref{fig:damping_pho+mag_ratio}. The two systems with sizable moments (Rh and Pt), clearly have a dominant contribution from spin fluctuations when the moments are reduced upon increased scattering due to temperature. In contrast, the two systems with (small) antiferromagnetic moments (Mo and W), the effect of the spin fluctuations on the damping is negligible and atomic displacements are solely responsible. The weak temperature dependence found in these doped Py systems is somewhat surprising since in pure metals like Fe and Ni, a strong temperature dependence has been both measured and calculated \cite{,Mankovsky2013}, however data for other random alloy systems is scarce. 

The temperature dependence of damping from the band structure is often attributed to interband and intraband transitions which arises from the torque-correlation model. Intraband transitions has conductivity like dependence on temperature while interband shows resistivty-like dependence. The weak overall dependence found in the systems in Fig.~\ref{fig:damping_pho+mag_ratio} suggests lack of intraband transitions but a more detailed analysis of the band structure and thermal disorder are left for a future study.



\subsection{Spin-wave stiffness and exchange stiffness}

\begin{figure}[htbp]
\begin{center}
\includegraphics[width=3.4in]{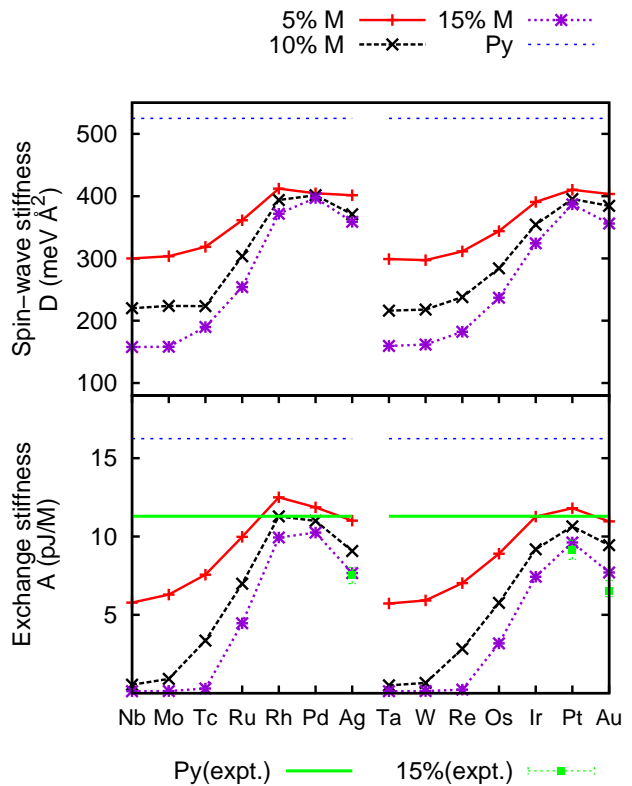}
\end{center}
\caption{\label{fig:stiffness_D_A}
Spin-wave stiffness $D$ of Py-M in the ground state (top) and exchange stiffness constant $A$ at room temperature $T=300\text{K}$ (bottom) as a function of doping concentration. The strict dashed lines show the reference value of pure Py from calculation and experiments. The scattered dots indicate the experimental data for Py+15\%M (Ag/Pt/Au)  from Ref.~\cite{tunable_yuli}
}
\end{figure}

 In the previous sections, we investigated saturation magnetization and damping and we are therefore left with the exchange stiffness. The calculated spin-wave stiffness $D$ at $T=0$ K, from Eq.~\ref{eq:D_origin}, is displayed in the upper panel of Fig.~\ref{fig:stiffness_D_A}. $D$ can be directly measured from neutron scattering experiments but as far as we are aware, no such data exist. For the late elements in the $4d$ and $5d$ series, the spin wave stiffness is maximized and have values rather similar to pure Py, however with a reduction of approximately 20\%. In micromagnetic modelling, it is common to use the exchange stiffness $A$ instead of $D$. $A$ is proportional to $D$, from Eq.~\ref{spin_stiff_A}, and the sole temperature dependence of $A$ therefore comes from the magnetization. In the lower panel of Fig.~\ref{fig:stiffness_D_A}, we show the calculated room temperature ($T=300$ K) values of $A$, together with values for pure Py and available experimental data. In the beginning of the $4d$ ($5d$) series, the exchange stiffness becomes small upon increasing concentration of impurities and the systems are magnetically very soft. It follows from the fact that magnetization is small because the systems are close to their ordering temperature. Contrary, for the late elements in the $4d$ ($5d$) series, the magnetization has a large finite value even at room temperature and therefore the exchange stiffness also has a large value, however reduced by approximately 15\% compared to pure Py.


\section{Summary and conclusions} \label{sec:summary}

A systematic study of the intrinsic magnetic properties of transition metal doped Py has been presented. It is found that the Gilbert damping is strongly dependent on the spin-orbit coupling of the impurity atoms and more weakly dependent on the density of states that determines disorder scattering. The strong influence of the spin-orbit coupling makes the $5d$ elements much more effective to change the Gilbert damping and more sensible to the concentration. As a result, the damping can be increased by an order of magnitude compared to undoped Py. Overall, the damping features are qualitatively rather well explained by the torque correlation model, yet it misses some quantitative predictive power that only first principles results can provide. Moreover, it is found that the damping overall has a weak temperature dependence, however it is slightly enhanced with temperature due to increased scattering caused by atomic displacements and spin fluctuations. Elements in the beginning of the $4d$ or $5d$ series are found to strongly influence the magnetization and exchange stiffness due to antiferromagnetic coupling between impurity and host atoms. In contrast, elements in the end of the $4d$ or $5d$ series keep the magnetization and exchange stiffness rather similar to undoped Py. More specifically, doping of the $5d$ elements Os, Ir and Pt are found to be excellent candidates for influencing the magnetodynamical properties of Py. 

Recently, there have been an increasing attention for finding metallic materials with small intrinsic damping, for instance half metallic Heusler materials and FeCo alloys \cite{Phillip,FeCodamping}. Controlling and varying the magnetodynamical properties in these systems through doping or by other means, like defects, are very relevant and left for a future study.  


\begin{acknowledgments}
The work was financed through VR (the Swedish Research Council) and GGS (G\"oran Gustafssons Foundation). A.B. acknowledge support from eSSENCE. The computations were performed on resources provided by SNIC (Swedish National Infrastructure for Computing) at NSC (National Supercomputer Centre) in Link\"oping.

\end{acknowledgments}

\bibliography{Gilbert_FP}
\end{document}